\documentclass[useAMS,usenatbib]{mn2e}
\usepackage{graphicx,epsfig,longtable}
\usepackage{color}
\usepackage{amsmath}
\usepackage{tabularx} 
\usepackage{hyperref}

\newcommand{\beqy}{\begin{eqnarray}}
\newcommand{\eeqy}{\end{eqnarray}}
\newcommand{\bmlet}{\begin{subequations}}
\newcommand{\emlet}{\end{subequations}}

\def\gsimeq{\,\,\raise0.14em\hbox{$>$}\kern-0.76em\lower0.28em\hbox
{$\sim$}\,\,}
\def\lsimeq{\,\,\raise0.14em\hbox{$<$}\kern-0.76em\lower0.28em\hbox
{$\sim$}\,\,}
\newcommand{\Msun}{$M_{\odot}$}
\newcommand{\aap}{A\&A}
\newcommand{\apjl}{ApJL}

\newcommand{\apj}{ApJ}

\newcommand{\prc}{Phys. Rev. C}
\newcommand{\prl}{Phys. Rev. Lett.}
\newcommand{\mnras}{MNRAS}

\newcommand{\physrep}{Phys. Rep.}
\newcommand{\prd}{Phys. Rev. D}

\title[Weak interactions in neutron-star mergers]
 {Impact of weak interactions of free nucleons on the r-process in dynamical ejecta from neutron-star mergers}

\author[S. Goriely et al.]{S.~Goriely$^1$, A.~Bauswein$^2$, O.~Just$^{3,5}$, E.~Pllumbi$^{3,4}$, and H.-Th.~Janka$^3$ \\
  $^1$Institut d'Astronomie et d'Astrophysique, Universit\'e Libre de Bruxelles,  CP 226, 1050 Brussels, Belgium \\
  $^2$Department of Physics, Aristotle University of Thessaloniki, 54124 Thessaloniki, Greece \\
  $^3$Max-Planck-Institut f\"ur Astrophysik, Postfach 1317, 85741 Garching, Germany \\
  $^4$Physik Department, Technische Universit\"at M\"unchen, James-Franck-Stra\ss e 1, 85748 Garching, Germany \\
  $^5$Max-Planck/Princeton Center for Plasma Physics (MPPC)
}

\date{Released 2015 Xxxxx XX}

\pagerange{\pageref{firstpage}--\pageref{lastpage}} \pubyear{2015}

\begin{document}
\label{firstpage}
\maketitle

\begin{abstract}
We investigate $\beta$-interactions of free nucleons and their impact on the 
electron fraction ($Y_e$) and r-process nucleosynthesis in ejecta characteristic
of binary neutron star mergers (BNSMs). For that
we employ trajectories from a relativistic BNSM model to represent the density-temperature
evolutions in our parametric study. In the high-density environment, positron captures
decrease the neutron richness at the high temperatures predicted by the
hydrodynamic simulation. Circumventing the complexities of modelling
three-dimensional neutrino transport, (anti)neutrino captures are parameterized in
terms of prescribed neutrino luminosities and mean energies, guided by published results
and assumed as constant in time. Depending sensitively on the adopted
$\nu_e$-$\bar\nu_e$ luminosity ratio, neutrino processes
increase $Y_e$ to values between 0.25 and 0.40, still allowing for a successful r-process
compatible with the observed solar abundance distribution and a significant
fraction of the ejecta consisting of r-process nuclei. If the $\nu_e$ luminosities and mean
energies are relatively large compared to the $\bar\nu_e$ properties, the mean $Y_e$ 
might reach values $>$0.40 so that neutrino captures seriously compromise the success
of the r-process. In this case, the r-abundances remain compatible with the solar
distribution, but the total amount of ejected r-material is reduced to a few percent,
because the production of iron-peak elements is favored. 
Proper neutrino physics, in particular also neutrino absorption, have to be included in BNSM
simulations before final conclusions can be drawn concerning r-processing in this environment
and concerning observational consequences like kilonovae, whose peak brightness and color 
temperature are sensitive to the composition-dependent opacity of the ejecta.
\end{abstract}


\section{Introduction}
\label{sect_intro}

The r-process, or rapid neutron-capture process, of stellar nucleosynthesis is invoked to explain
the production of the stable (and some long-lived radioactive) neutron-rich nuclides heavier than
iron that are observed in stars of various metallicities, as well as in the solar system \citep[for
a review, see ][]{arnould07}.  Despite important effort to model potential r-process sites, all
the proposed scenarios face serious problems and the site(s) of the r-process is (are) not
identified yet.  Until now, type-II supernovae or $\gamma$-ray bursts models have failed to provide
convincing evidence for a successful r-processing that could significantly contribute to the
Galactic enrichment in r-material \citep{wanajo11,janka12,burrows13}.  Only magneto-rotational
supernova explosions with extremely strong pre-collapse magnetic fields and fast rotation seem to provide
favourable conditions for r-processing, but such rare events are not expected to be at the origin of
the global galactic enrichment in r-process nuclei \citep{winteler12,nishimura15,wehmeyer15}.

For this reason, special attention is now being paid to neutron star (NS) mergers following the
confirmation by hydrodynamic simulations that a significant amount of r-process enriched material,
typically about $10^{-3}$ to a few $10^{-2}$\Msun, can be ejected \citep{rosswog99,frei99,arnould07,metzger10,roberts11,goriely11,korobkin12,bauswein13,goriely13,wanajo14,perego14,just15,seki15}.
Recent nucleosynthesis calculations by \citet{just15} show that the combined
contribution of both the dynamical (prompt) ejecta expelled during binary NS or NS-black hole
(BH) mergers and the neutrino and viscously driven outflows generated during the post-merger
remnant evolution of relic BH-torus systems can lead to the production of r-process elements from
mass number $A \ga90$ up to thorium and uranium. The corresponding abundance distribution reproduces 
the solar distribution extremely well and can also account for 
the elemental distributions observed in low-metallicity stars
\citep{roederer11,roederer12}.  Furthermore, recent studies
\citep{mat14,kom14,mennekens14,shen14,voort14,vangioni14,wehmeyer15} have reconsidered the galactic
or cosmic chemical evolution of r-process elements in different evolutionary contexts. Although
they do not converge towards one unique quantitative picture, most of them arrived at the conclusion
that double compact star mergers may be the major production sites of r-process elements.

Despite the recent success of nucleosynthesis studies for NS mergers, the possibility of r-processing
in these events is still affected by a variety of uncertainties.
In particular, the impact of neutrino interactions is not yet studied and understood in detail,
the main reason of which is the not yet manageable computational complexity associated with 
neutrino transport in a generically three-dimensional (3D), highly asymmetric environment with nearly
relativistic fluid velocities and rapid changes in time. A computationally simple and much more 
efficient alternative to solving the time-dependent transport equation for neutrino distributions 
in the six-dimensional phase space is the use of a neutrino 
leakage scheme, in which the local neutrino net-emission (i.e., emission minus absorption) rate is
estimated by a weighted interpolation between the pure emission rate and an optical-depth dependent 
diffusive loss term \citep[e.g.,][]{Ruffert1996a, Rosswog2003}.
However, neutrino absorption cannot be straightforwardly included in a
self-consistent manner in a leakage scheme because information about the local neutrino
densities is missing in such a treatment.

It was long believed that neutrino interactions could not, at least not drastically, affect the
initial neutron richness of the ejecta. Such expectations were based on numerical merger 
models. The temperatures of the merging NSs in these simulations, and therefore the
neutrino production rates, remain low until the NSs collide with each other, and they rise in an 
increasing volume only gradually on a time scale of several milliseconds after the first contact of
the two NSs \citep{Ruffert1999a,Rosswog2003,wanajo14}. At this time a significant fraction
of the ejecta material is already being expelled with nearly relativistic velocities. 
Newtonian merger
studies suggested that a large part of the ejecta material even from symmetric mergers (i.e., for
two equal-mass or close to equal-mass NSs) is thrown out by tidal forces in extended spiral
arms forming from matter of the outer faces of the merging objects during their final approach 
and collision \citep{rosswog99,korobkin12}. By the tidal stretching these arms naturally
remain unshocked and thus stay cool. Therefore they reach low densities so quickly that
electron and positron captures cannot become efficient. Moreover, the ejecta escape to large
radii before the neutrino emission of the compact merger remnant becomes sizable and 
neutrino absorption can affect the electron fraction significantly. 
Absorption of neutrinos radiated by the massive merger remnant is also diminished because
the tidal ejection in Newtonian models happens preferentially in the orbital plane \citep[see
e.g.][]{rosswog99,korobkin12} while the neutrinos are predominantly emitted perpendicular
to this plane \citep{Rosswog2003, Dessart2009, perego14}. However, the described
situation applies well only for Newtonian mergers. The situation is different in the 
relativistic case.

Relativistic simulations of symmetric mergers do not show the development of prominent
tidal arms, and the ejection of unshocked matter is therefore not important. Instead,
the collision shock that builds up at the interface of the two NSs is typically stronger 
than in Newtonian conditions \citep[see e.g.][]{bauswein13}, leading to
potentially higher temperatures and higher, faster rising neutrino luminosities 
\citep{wanajo14, seki15}. The ejecta are expelled fairly spherically instead of equatorially
\citep{bauswein13,Hotokezaka2013} and consist of two main components, 
namely a first one in which
very hot material is squeezed out from the collision interface of the two merging bodies,
and a second, slightly delayed one that is expelled in waves from a torus-like belt of matter
around the high-density core of the merger remnant. This torus is heated by spiral shocks 
that are sent out from the aspherical, wobbling, and rotating high-density core and that
also lead to outward acceleration of parts of the torus matter
\citep{bauswein13,Hotokezaka2013,just15,wanajo14,seki15}.

Indeed, recent relativistic NS-NS merger simulations that took into account
neutrino emission by means of a leakage scheme and absorption by an additional
approximate transport treatment based on a moment formalism \citep{wanajo14,seki15}, 
found that neutrino interactions with free nucleons can 
significantly increase the electron fraction in the dynamical ejecta for cases in which
the collapse of the merger remnant to a black hole is delayed or does not happen.
Under such conditions nuclei with mass numbers $A<140$ can also be created in the dynamical 
ejecta in addition to the heavy r-process elements ($A>140$). Weak interaction
processes of free nucleons can consequently affect the strength of the r-process
and the emerging abundance distribution.

The accurate inclusion of neutrino interactions in hydrodynamical simulations 
remains a highly complex task. This motivated us to conduct a simple, parametric
study in order to quantify the potential impact of weak interactions on the
electron-fraction evolution in merger ejecta and thus to explore the consequences
of charged-current neutrino-nucleon reactions for the nucleosynthesis and
possible r-processing in these ejecta. More specifically, we investigate the 
influence of $\beta$-interactions of electron neutrinos ($\nu_e$) and electron
antineutrinos ($\bar{\nu}_e$) with free $n$ and $p$ and of their inverse reactions,
\begin{eqnarray}
& & \nu_e+ n \rightleftharpoons p + e^- \label{eq:betareac1}\\
& & \bar{\nu}_e + p \rightleftharpoons n + e^+ \quad \,, 
\label{eq:betareac2}
\end{eqnarray}
on the $Y_e$ distribution and r-process nucleosynthesis at conditions representative
of the dynamical ejecta expelled by hydrodynamical forces during NS-NS mergers.
These reactions have been neglected in all previous studies of r-process 
nucleosynthesis for such ejecta except those of \citet{wanajo14},
where their quantitative effects may depend 
on the adopted equation of state \citep{seki15}. The role of 
weak interactions for the electron fraction and the
corresponding implications for r-process nucleosynthesis, however,
demand further exploration in more detail, in particular also by basic, parametric
modeling, because a multitude of uncertainties will prevent rigorous, 
self-consistent solutions of the full problem 
in the near future. Such uncertainties are associated with, for example, the extreme
complexities of 3D energy-dependent neutrino transport in relativistic environments,
with the neutrino opacities of dense, potentially highly magnetized matter, and with
neutrino-flavor oscillations at rapidly time-variable, largely aspherical
conditions of neutrino emission.

The main objective of our present work is a sensitivity study by the use of
a parametric approach. It is intended to motivate further explorations 
of neutrino effects in relativistic NS-NS mergers in more detail and breadth.
To this end, we set up a simplified and idealized theoretical framework 
to test the individual roles of the different weak interaction processes,
considering the density and temperature evolution of fluid elements ejected 
from a prototype hydrodynamical, relativistic NS-NS merger model.
For our parameter study we make assumptions about the neutrino emission properties
that are guided by data taken from the literature. 
The electron fractions resulting from the neutrino-processing of the ejecta elements
are then used as input for nuclear network calculations, allowing us to immediately
link the effects of neutrino interactions to the final heavy-element production.

In Sect.~\ref{sect_weak} the employed merger model and our
treatment of the weak neutrino reactions with free nucleons are
described. The effects of $\beta$-processes on the electron fraction are reported in
Sect.~\ref{sect_ye}, and the subsequent r-process nucleosynthesis is analysed in
Sect.~\ref{sect_rpro}. Conclusions are drawn in Sect.~\ref{sect_conc}.

\section{Weak interactions of free nucleons in merger ejecta}
\label{sect_weak}

We adopt the 
density evolution of ejecta fluid elements from a representative NS-NS merger model, namely the
symmetric 1.35\Msun-1.35\Msun\ binary model obtained with the temperature-dependent DD2 equation
of state \citep{hempel10,typel10} \citep[essentially identical to the one in][but with a higher
resolution of $\sim$10$^6$ particles]{bauswein13}. This relativistic hydrodynamic simulation
also provides the temperature evolution. We do not apply any temperature post-processing
as in \citet{goriely11} to disentangle temperature jumps in shocks from artificial heating
associated with the use of a numerical viscosity in the smoothed-particle hydrodynamics
scheme.  

For each trajectory, we follow the expansion starting at a fiducial density of
$\rho_\mathrm{eq}=10^{12}$\,g\,cm$^{-3}$,
where we assume equilibrium to hold between electrons,
positrons and neutrinos for the given total lepton number provided by our 
NS-NS merger model. Below $\rho_\mathrm{eq}=10^{12}$\,g\,cm$^{-3}$,
electron, positron and electron neutrino and antineutrino captures are
systematically included. Reactions of neutrinos on nuclei are, however, neglected. As long as
the temperature remains in excess of typically $T>10^{10}$~K, the abundance of heavy nuclei
is determined by nuclear statistical equilibrium (NSE) at the given electron fraction, density and
temperature. From the density $\rho_\mathrm{eq}$ down to the density $\rho_\mathrm{net}$,
at which the temperature reaches 10\,GK and the full reaction network is initiated, the 
$\beta$-interactions of free nucleons may affect the electron fraction $Y_e$. If a trajectory
stays cooler than 10\,GK below the density $\rho_\mathrm{eq}$, the network calculation is
started at the neutron-drip density $\rho_\mathrm{drip}$, i.e.,
$\rho_\mathrm{net}=\rho_\mathrm{drip}\simeq 4.2\times 10^{11}$\,g\,cm$^{-3}$. 
The considered $\beta$-reactions involve free nucleons, whose
abundance variations are given by
\begin{eqnarray}
\frac{dY_n^\mathrm{f}}{dt}&=& -\lambda_+ Y_n^\mathrm{f} + \lambda_- Y_p^\mathrm{f} \nonumber\,, \\ 
\frac{dY_p^\mathrm{f}}{dt}&=&\lambda_+ Y_n^\mathrm{f} - \lambda_- Y_p^\mathrm{f} \,,
\label{eq_dyq}
\end{eqnarray}
where $\lambda_+=\lambda_{\nu_e} + \lambda_{e^+}$ and
$\lambda_- =\lambda_{{\bar\nu}_e} + \lambda_{e^-}$. The $\lambda_x$ denote capture rates of
species $x\in\{e^-,e^+,\nu_e,\bar\nu_e\}$ onto free nucleons according to the $\beta$-reactions,
Eqs.~(\ref{eq:betareac1},\ref{eq:betareac2}). The free neutron and proton numbers are related to
$Y_e$ by
\begin{eqnarray}
Y_n^\mathrm{f}&=& 1-Y_e-\sum_{Z \ge 2} N~Y(Z,N)\nonumber\,, \\ 
Y_p^\mathrm{f}&=& Y_e-\sum_{Z \ge 2} Z~Y(Z,N)\,,
\label{eq_yf}
\end{eqnarray}
where $Y=X/A$ is the molar fraction (and $X$ the mass fraction) of the nucleus ($Z,N$) of atomic
mass $A=Z+N$. Assuming that the NSE molar fractions of nuclei remain constant over the time step
$\Delta t$, the time evolution of $Y_e$ can be related to the change of the number of free protons,
and written as
\begin{equation}
\frac{dY_e}{dt}=-\lambda_{\rm tot}Y_e + \lambda^* \,,
\label{eq_dye}
\end{equation}
where $\lambda_{\rm tot}=\lambda_+ + \lambda_-$ and 
\begin{equation}
\lambda^*=\lambda_+ \left[ 1-\sum_{Z\ge 2}NY \right]+\lambda_- \sum_{Z\ge 2}ZY \,.
\label{eq_ls}
\end{equation}
If only free neutrons and protons are present, $\lambda^*=\lambda_+$. The impact of heavy nuclei is
to increase $\lambda^*$ towards $ \lambda_-$ (for matter made of $\alpha$-particles only,
$\lambda^*=\lambda_{\rm tot}/2$). The $\alpha$-effect \citep{mclaughlin96,meyer98,pllumbi14}, or
more generally, the effect of heavy nuclei in binding neutrons and protons inside nuclei, is
known to be responsible for driving $Y_e$ towards 0.5
and is included in this term $\lambda^*$. If we assume that $\lambda_{\rm tot}$ and
$\lambda^*$ remain constant over the time step $\Delta t$ (or, specifically, that $\Delta t$ is chosen
such that $\lambda_{\rm tot}$ and $\lambda^*$ as well as the abundance of nuclei remain essentially
constant during the time step), Eq.~(\ref{eq_dye}) can be integrated analytically leading to
\begin{equation} Y_e(t+\Delta t)\simeq Y_e(t) e^{(-\lambda_{\rm tot} \Delta t)} +
\frac{\lambda^*}{\lambda_{\rm tot}} \left[ 1-e^{(-\lambda_{\rm tot} \Delta t)} \right] \,,
\label{eq_ye}
\end{equation} where $\lambda_{\rm tot}$ and $\lambda^*$ are estimated at time $t$. This equation is
used to follow $Y_e$ from the initial density $\rho_\mathrm{eq}$ down to the density $\rho_{\rm net}$,
at which the temperature has dropped to 10~GK (or, alternatively, to the drip density if
temperatures above 10~GK are not reached for the considered trajectory).

For $t \gg 1/\lambda_{\rm tot}$, $Y_e$ reaches the equilibrium value
$Y_e^\infty$ given by
\begin{equation} Y_e^\infty\simeq \frac{\lambda^*}{\lambda_++\lambda_-} \simeq
  \frac{\lambda_{\nu_e} + \lambda_{e^+}}{\lambda_{\nu_e} + \lambda_{e^+}+\lambda_{{\bar\nu}_e} +
    \lambda_{e^-}} \quad .
\label{eq_yeinf}
\end{equation} 
The electron (anti)neutrino capture rates can be written in terms of the average
(anti)neutrino capture cross sections $\langle \sigma_{\nu_e/\bar\nu_e} \rangle$ \citep{pllumbi14}
as
\begin{eqnarray} \lambda_{\nu_e} & \simeq & \frac{L_{\nu_e}}{4\pi r^2 \langle E_{\nu_e}\rangle}
\langle \sigma_{\nu_e} \rangle \,,
\label{eq_lnu}\\ \lambda_{{\bar\nu}_e} & \simeq & \frac{L_{{\bar\nu}_e}}{4\pi r^2 \langle
E_{{\bar\nu}_e}\rangle} \langle \sigma_{{\bar\nu}_e} \rangle \,.
\label{eq_lnub}
\end{eqnarray}
Here, the local (anti)neutrino number densities are expressed by the ratios of the global 
luminosities, $L_{\nu_e/\bar\nu_e}$, and the mean energies of the radiated neutrinos,
$\langle E_{\nu_e/\bar\nu_e}\rangle$, multiplied with the spherical surface $4\pi r^2$ that
surrounds the central neutrino source at a radial distance $r$ (for every trajectory we adopt the time-dependent radial distance $r$ from our hydrodynamical model). This $r^{-2}$ dilution of the neutrino flux is a crude approximation 
and holds, at best, far away from the neutrinosphere, provided the emission is isotropic, i.e.,
if directional variations of the neutrino fluxes do not play a role. Close to and below the 
neutrinosphere, however, such a description breaks down but can be justified by the
fact that at these locations electron and positron captures dominate and their competition
enforces a state of weak equilibrium. At large distances the asymptotic electron fraction
is determined by neutrino and antineutrino absorptions, for which reason
direction-dependent differences of
the neutrino exposure of the ejecta would be important for a detailed discussion of neutrino
effects on merger ejecta. Nevertheless, despite these shortcomings, we apply the simple 
ansatz at all radii $r$ where $\rho\le\rho_\mathrm{eq}$ in order to discuss basic aspects
of the impact of neutrino processes with nucleons in the merger ejecta in a parametric way.

Similar to Eqs.~(\ref{eq_lnu},\ref{eq_lnub}), the electron and positron capture rates are given 
in terms of the 
average electron/positron capture cross sections $\langle \sigma_{e^-/e^+} \rangle$ by
\begin{eqnarray} 
\lambda_{e^+} & = &c~ {\tilde n_{e^+}} \langle \sigma_{e^+} \rangle \,, \\ \lambda_{e^-} & = &c ~
n_{e^-} \langle \sigma_{e^-} \rangle \,,
\label{eq_lep}
\end{eqnarray} 
where $c$ is the speed of light and $n_{e^-}$ and ${\tilde n_{e^+}}$ the electron and
positron densities, as detailed in \citet{pllumbi14}.

In turn, the average cross sections for electron neutrino and antineutrino captures, as well as those
for electron and positron captures, including the weak magnetism and recoil corrections, are taken from
\citet{pllumbi14} \citep[see also][]{horowitz99}. While the electron and positron capture rates are
temperature- and density-dependent only, the (anti)neutrino capture rates depend on the
(anti)neutrino luminosities and mean energies, hence require a detailed knowledge of the neutrino
properties at each time step.

For the present study, we consider representative (anti)neutrino luminosities and
angle-averaged mean energies that are assumed to remain constant in time. For a given mean energy
$\langle E_{\nu_e}\rangle$, the electron neutrino temperature $T_{\nu_e}$ is deduced from the
relation 
\begin{equation} \langle E_{\nu_e}\rangle=k_\mathrm{B}T_{\nu_e} \cdot \frac{F_3(0)}{F_2(0)} \quad ,
\label{eq_tnu}
\end{equation} 
where $k_\mathrm{B}$ is the Boltzmann constant and $F_n$ are the fermi integrals \citep{takahashi78} 
of order $n$ for vanishing chemical potential, assuming nondegenerate neutrino spectra.
A similar expression holds for the anti\-neutrino temperature. For given (anti)neutrino luminosities
and mean energies this allows us to determine all other moments of the
(anti)neutrino energy spectra, and consequently the corresponding capture cross sections and rates
(Eqs.~\ref{eq_lnu},\ref{eq_lnub}).

\begin{figure}
\includegraphics[scale=0.3]{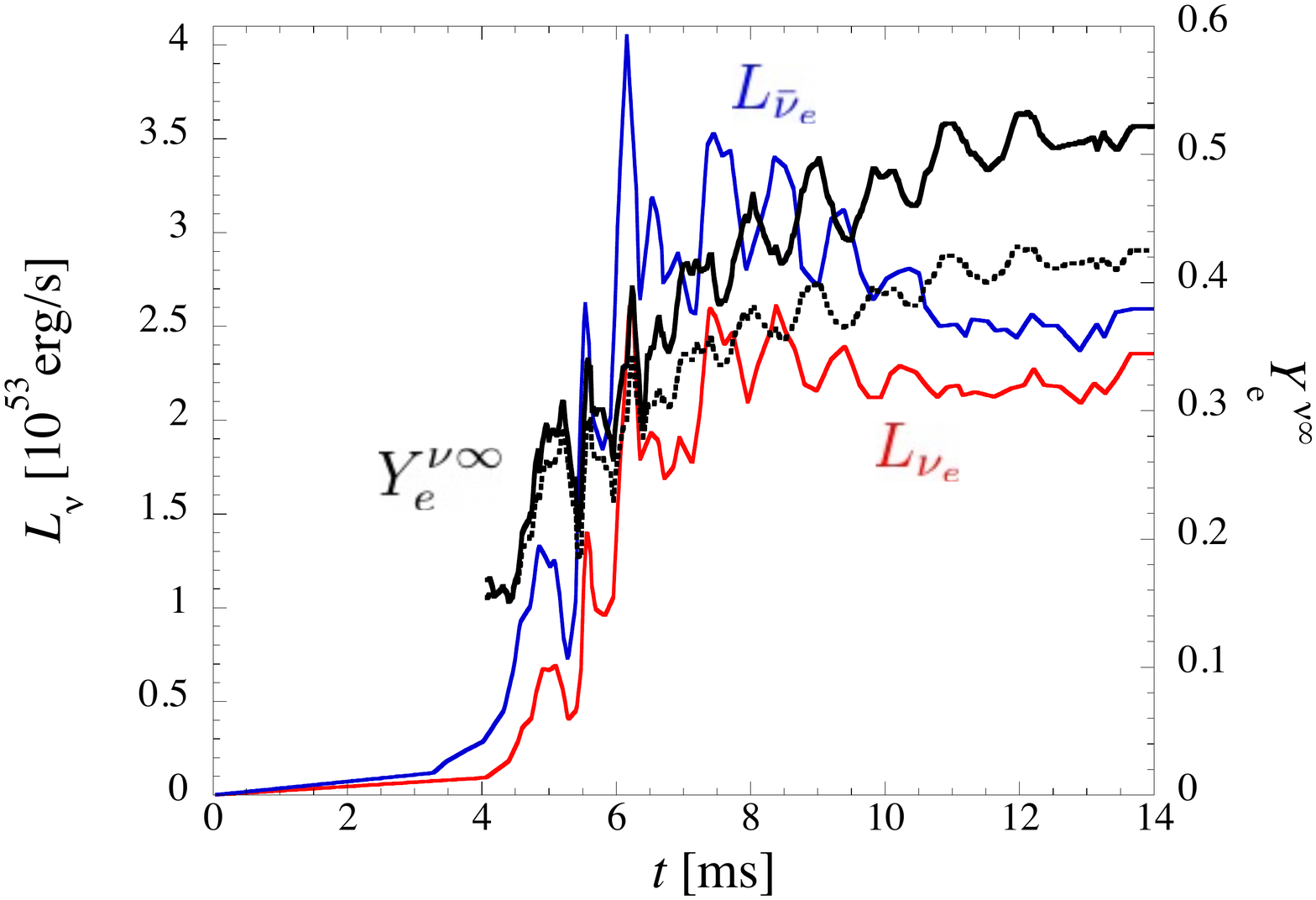}
\caption{(Color online). Time evolution of the electron neutrino and antineutrino luminosities from
\citet{wanajo14} and the corresponding $Y_e^{\nu\infty}$ (black solid line) as defined by
Eq.~(\ref{eq_yeinf_nu}). The black dotted line gives the $Y_e^{\nu\infty}$ without weak
magnetism and recoil corrections, $i.e.$ $f_\nu^\mathrm{mr}=1$ in Eq.~(\ref{eq_yeinf_nu}). The
(anti)neutrino mean energies are taken consistently from Fig.~1 (lower panel) of
~\citet{wanajo14}. Asymptotic values of $Y_e^{\nu\infty}$ are calculated only for non-negligible
neutrino luminosities, $i.e.$ for times $t\gsimeq 4$~ms.  }
\label{fig_yeinf1}
\end{figure} 

\begin{figure}
\includegraphics[scale=0.3]{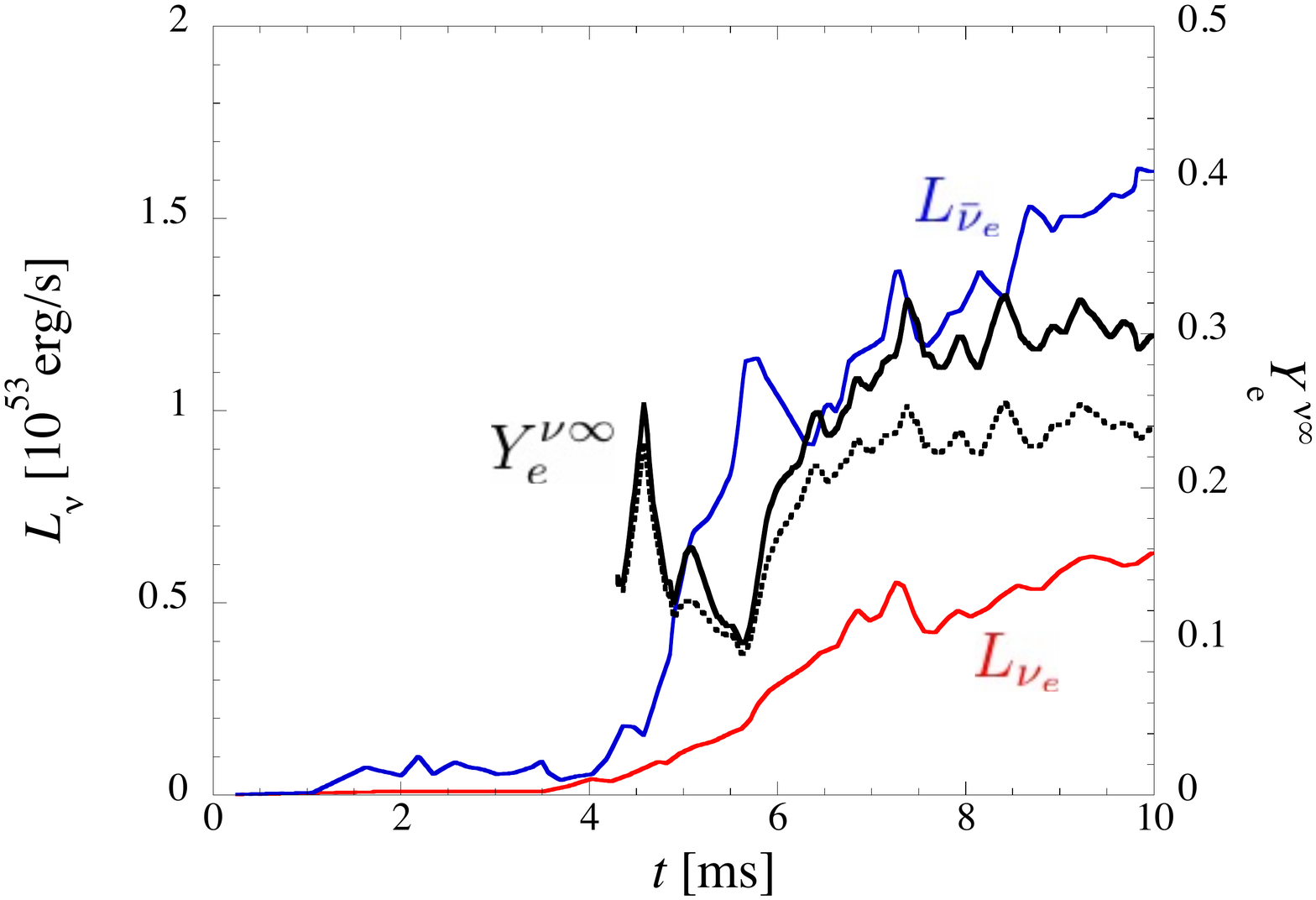}
\caption{(Color online). Same as Fig.~\ref{fig_yeinf1}, but for the electron neutrino and antineutrino
luminosities from \citet{ruffert01}. The (anti)neutrino mean energies are taken consistently from
Fig.~17 (lower panel; Model Bc) of this reference.}
\label{fig_yeinf2}
\end{figure} 

Assuming that electron (anti)neutrino captures dominate over electron and positron captures and
the abundance of heavy nuclei is negligible, the asymptotic value of $Y_e$ (Eq.~\ref{eq_yeinf}) can
be approximated by
\begin{equation} Y_{e}^{\nu\infty}\simeq \frac{L_{\nu_e} \varepsilon_{\nu_e} f_{\nu_e}^\mathrm{mr}}
{L_{\nu_e} \varepsilon_{\nu_e} f_{\nu_e}^\mathrm{mr} + L_{{\bar\nu}_e} \varepsilon_{{\bar\nu}_e}
f_{{\bar\nu}_e}^\mathrm{mr}} \,,
\label{eq_yeinf_nu}
\end{equation} 
because the rates can be expressed as $\lambda_{\nu}\propto L_{\nu}\varepsilon_{\nu}f_{\nu}^\mathrm{mr}$
($\nu=\nu_e, {\bar\nu}_e$), where
$ \varepsilon_{\nu}=\langle E^2_{\nu} \rangle / \langle E_{\nu}\rangle=F_4(0)F_2(0)/F_3^2(0) \times
\langle E_{\nu}\rangle$ and
$f_{\nu}^\mathrm{mr}$ corresponds to the weak magnetism and recoil corrections that can be found in
\citet{pllumbi14}.  The corresponding value of $Y_{e}^{\nu\infty}$ is shown in Fig.~\ref{fig_yeinf1}
for the luminosities and mean energies of \citet{wanajo14} and lies between 0.25 and 0.50 for
times $t\gsimeq 5$\,ms after the merging of the binary NSs. These values may, however, still
be modified by effects of heavy nuclei. The weak magnetism and recoil corrections on the (anti)neutrino
rates are seen in Fig.~\ref{fig_yeinf1} to increase the asymptotic value of $Y_{e}^{\nu\infty}$ by up
to 20\% because they reduce the antineutrino capture cross section and simultaneously increase the
neutrino capture cross section \citep{horowitz99}.

The neutrino properties, and in particular the antineutrino to neutrino luminosity ratio, are found
to vary significantly between different hydrodynamical simulations but also depend on the adopted
equation of state \citep{seki15}. Avoiding the complexity of self-consistent neutrino transport in
hydrodynamical simulations, we shall restrict ourselves in our sensitivity study to constant
luminosities and mean energies taken at selected times from previous simulations
\citep{ruffert01,wanajo14}. We consider first two representative sets of values for electron (anti)neutrino
luminosities and mean energies as obtained by \citet{wanajo14} (see their Fig.~1), namely those
corresponding to the instants of 5 and 6\,ms, i.e.,
\begin{itemize}
\item{Case 1:} $t\simeq 5$~ms with $L_{\nu_e} =0.6\times 10^{53}$~erg/s; $L_{{\bar\nu}_e}
=1.3\times 10^{53}$~erg/s; $\langle E_{\nu_e}\rangle=12$~MeV; $ \langle E_{{\bar\nu}_e}\rangle = 16$~MeV\,,
\item{Case 2:} $t\simeq 6$~ms with $L_{\nu_e} =2.6\times 10^{53}$~erg/s; $L_{{\bar\nu}_e}
=4.0\times 10^{53}$~erg/s; $\langle E_{\nu_e}\rangle=13$~MeV; $ \langle E_{{\bar\nu}_e}\rangle = 16$~MeV\,.
\end{itemize} 
Note that Case 1 leads to an asymptotic value $Y_{e}^{\nu\infty} \simeq 0.31$, whereas Case 2 yields
$Y_{e}^{\nu\infty}\simeq 0.42$ (Fig.~\ref{fig_yeinf1}).

In order to test more thoroughly the impact of the neutrino processes on the nucleosynthesis, we
also consider here the electron (anti)neutrino properties calculated in the NS merger simulation of
\citet{ruffert01} (cf Model Bc in their Fig.~17). The corresponding (anti)neutrino luminosities are
shown in Fig.~\ref{fig_yeinf2}, together with the asymptotic values $Y_{e}^{\nu\infty}$ (with and
without the weak magnetism and recoil corrections). Overall, lower luminosities are found in this
model in comparison with \citet{wanajo14}, but also a relatively higher emission of electron
antineutrinos. Consequently, lower 
values of $Y_{e}^{\nu\infty}$ are predicted for this model. In our present sensitivity analysis 
we also select two cases for the electron (anti)neutrino luminosities and mean energies from the
results of \citet{ruffert01}, namely those corresponding to times of 6 and 10\,ms (and hereafter 
referred to as Cases 3 and 4, respectively), i.e.,
\begin{itemize}
\item{Case 3:} $t\simeq 6$~ms with $L_{\nu_e} =0.3\times 10^{53}$~erg/s; $L_{{\bar\nu}_e} =10^{53}$~erg/s;
$\langle E_{\nu_e}\rangle=12.5$~MeV; $ \langle E_{{\bar\nu}_e}\rangle = 17.4$~MeV \,,
\item{Case 4:} $t\simeq 10$~ms with $L_{\nu_e} =0.6\times 10^{53}$~erg/s; $L_{{\bar\nu}_e}
=1.6\times 10^{53}$~erg/s; $\langle E_{\nu_e}\rangle=13.5$~MeV; $ \langle E_{{\bar\nu}_e}\rangle =
16.3$~MeV \,.
\end{itemize} 
Case 3 leads to an asymptotic value $Y_{e}^{\nu\infty} \simeq 0.21$, while Case 4 yields
$Y_{e}^{\nu\infty}\simeq 0.29$ (Fig.~\ref{fig_yeinf2}).

Finally, it should be mentioned that our assumption of time-independent
(anti)neutrino luminosities can be questioned, since the neutrino-ejecta interaction is a highly
time-dependent problem, where the relative time between the growth of the neutrino emission and the
mass ejection matters. Considering constant luminosities is a very crude but simple approximation,
which is sufficiently good to demonstrate the impact of neutrino processes on the time evolution and
mass distribution of the electron fraction and the corresponding consequences for the r-process.
A predictive assessment of neutrino effects on the nucleosynthesis in merger ejecta would also have 
to take account variations of the neutrino emission with different directions. Matter expelled 
towards the polar directions is exposed to different neutrino conditions than matter that leaves
the system along equatorial trajectories. Again, a more detailed description of neutrino transport
effects is demanded and is beyond the scope of our present parametric study.

\section{Impact of $\beta$-interactions on the electron fraction}
\label{sect_ye}

\subsection{Time evolution of $Y_e$}

\begin{figure}
\includegraphics[scale=0.4]{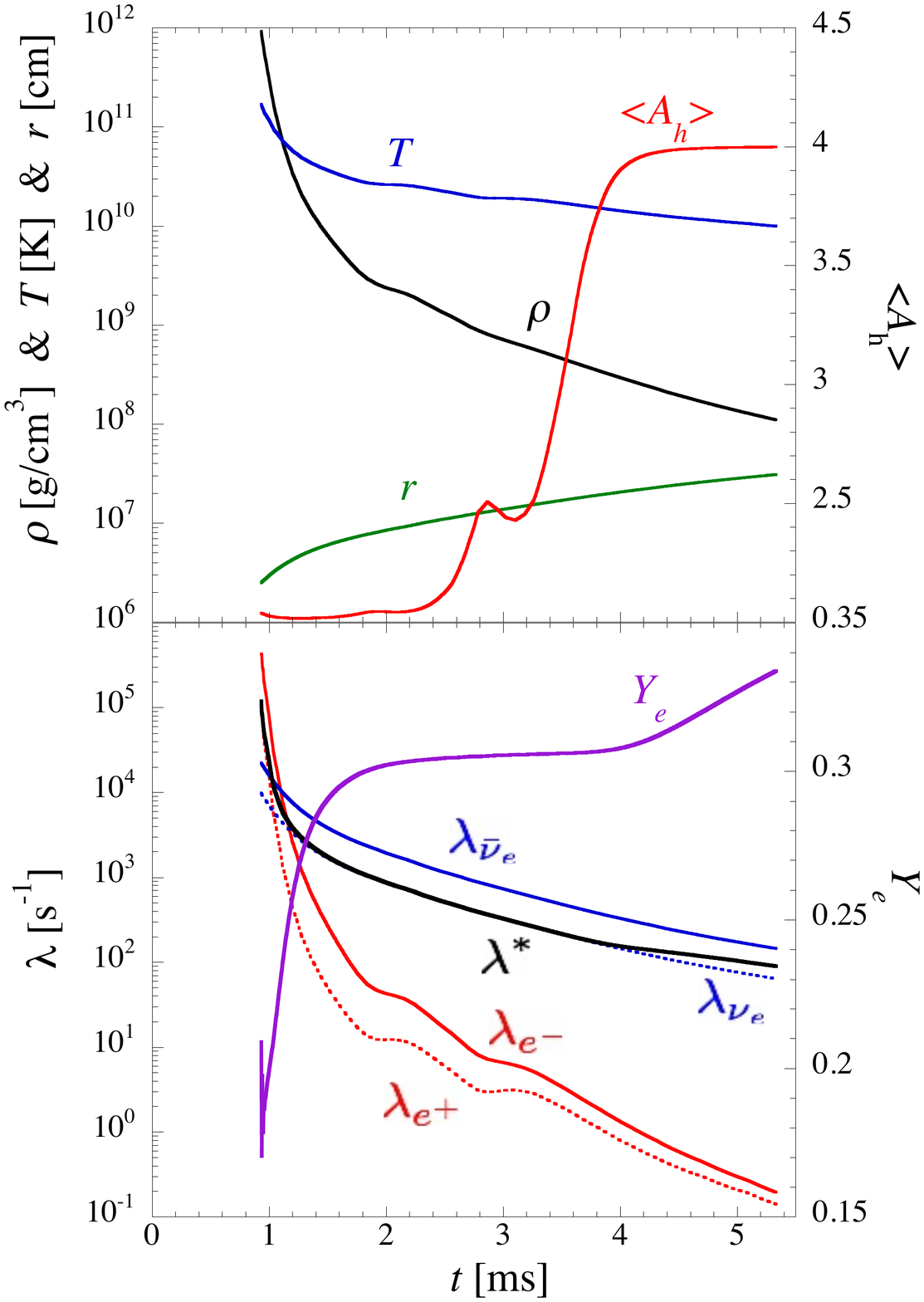}
  \caption{(Color online). {\it Upper panel}: Time evolution of temperature, density, radius,
   and mean atomic mass of nuclei heavier than protons, $\langle A_h\rangle=\sum_{Z\ge 2}AY/\sum_{Z\ge 2}Y$,
    between $\rho_\mathrm{eq}$ and $\rho_{\rm net}$ for trajectory 400720.
    {\it Lower panel}: 
    Analogue for the electron (anti)neutrino, electron and positron capture rates, and for
    $\lambda^*$ (Eq.~\ref{eq_ls}) and the electron fraction. The (anti)neutrino properties
    correspond to Case~1. The late-time increase of $Y_e$ is caused by the $\alpha$-effect,
    which does not asymptote to a terminal value until the network is started. $Y_e$ continues
    to evolve subsequently during the nucleosynthesis because of $\beta$-decays.
      }
\label{fig_traj1}
\end{figure}

\begin{figure}
\includegraphics[scale=0.4]{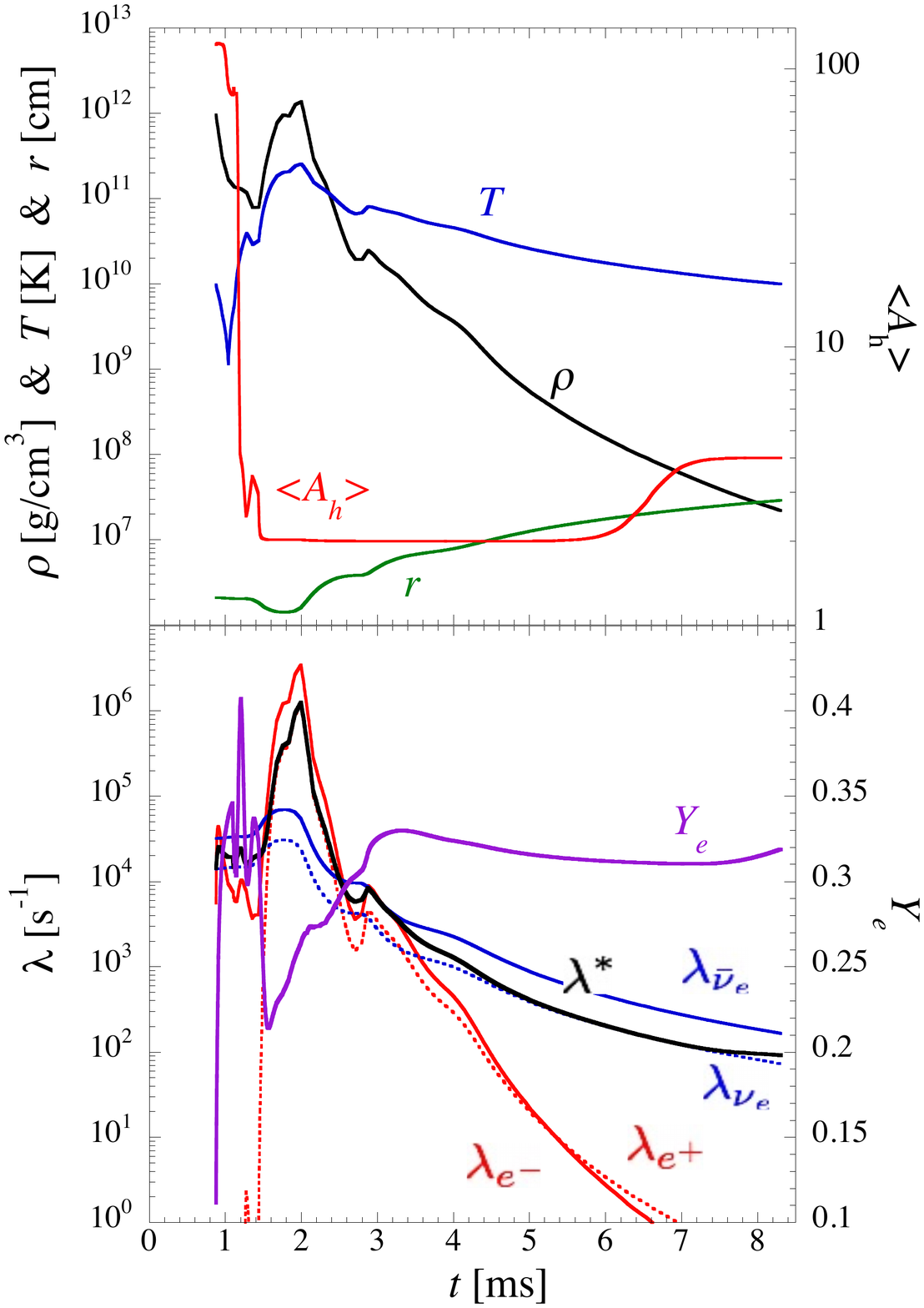}
  \caption{(Color online). Same as Fig.~\ref{fig_traj1}, but for trajectory 486857.}
\label{fig_traj2}
\end{figure}

To illustrate the impact of electron (anti)neutrino, electron and positron captures on the
evolution of $Y_e$, we show in Figs.~\ref{fig_traj1} and \ref{fig_traj2} the time evolution for two
specific trajectories during the expansion from the initial density 
$\rho_\mathrm{eq}=10^{12}$\,g\,cm$^{-3}$
down to density $\rho_{\rm net}$, where the reaction network calculations are initiated. Both
trajectories are studied including (anti)neutrino captures with neutrino properties for Case~1. 


For trajectory 400720 (Fig.~\ref{fig_traj1}), neutrino and antineutrino captures on free nucleons
with rates up to some $10^4~{\rm s^{-1}}$ lead to a rapid increase of $Y_e$ from
0.17 up to 0.31, corresponding to the asymptotic value of $Y_{e}^{\nu\infty}$ in Case~1. The
(anti)neutrino rates are found to dominate the electron and positron capture rates already
at densities slightly below $\rho_\mathrm{eq}$. This is linked to the slow $1/r^2$ decrease of
the (anti)neutrino fluxes, which is more shallow than the steep temperature dependence 
(roughly like $T^6$) of the electron and positron capture rates. At $t > 4$\,ms, free neutrons and
protons in the expanding matter partially recombine into $\alpha$-particles
($\langle A_h \rangle=\sum_{Z\ge 2} AY / \sum_{Z\ge 2} Y\simeq 4$), thus giving rise to the 
$\alpha$-effect and therefore a further increase of $Y_e$. The rate $\lambda^*$ becomes larger than
$\lambda_+\simeq \lambda_{\nu_e} $ and approaches $(\lambda_{{\bar\nu}_e}+\lambda_{{\nu}_e})/2$ of
material dominated by $\alpha$-particles.

For trajectory 486857 (Fig.~\ref{fig_traj2}), at the initial density $\rho_\mathrm{eq}$, the mass element
is characterised by a low temperature $T < 10^{10}$~K and consequently is
composed of heavy nuclei $\langle A_h \rangle \simeq 120$ typical of the low values of 
$Y_e\simeq 0.1$ in the 
outer NS crust. The fast $e^-$ and (anti)neutrino capture rates lead to a rapid increase of $Y_e$,
but all these rates are comparable and $Y_e$ fluctuates wildly. The
mass element is then subject to a new compression phase that sets in 
at $t\simeq 1.5$~ms, and the corresponding high
temperatures ($T\gsimeq 10^{11}$~K) photodissociate the matter into free nucleons. It should be noted
that during this recompression episode, the higher densities cause neutrinos to become trapped again, 
for which reason the application of Eqs.~(\ref{eq_lnu},\ref{eq_lnub}) remains problematic and
highly schematic. However, at these high-density, high-temperature conditions, electron captures
dominate and re-neutronize the material until they are counterbalanced by positron captures, whose rate
increases dramatically to achieve weak equilibrium with the electron captures. During the subsequent
expansion phase (at $t>2$\,ms), $Y_e$ rises gradually until (anti)neutrino absorptions take over 
to push $Y_e$ towards its asymptotic value of $Y_{e}^{\nu\infty}\simeq 0.31$ (at $t\sim$6\,ms). With
decreasing temperatures, the $\alpha$-effect finally becomes responsible for a late increase of
$Y_e$ at $t \gsimeq 7.5$\,ms.

\begin{figure*}
\includegraphics[scale=0.55]{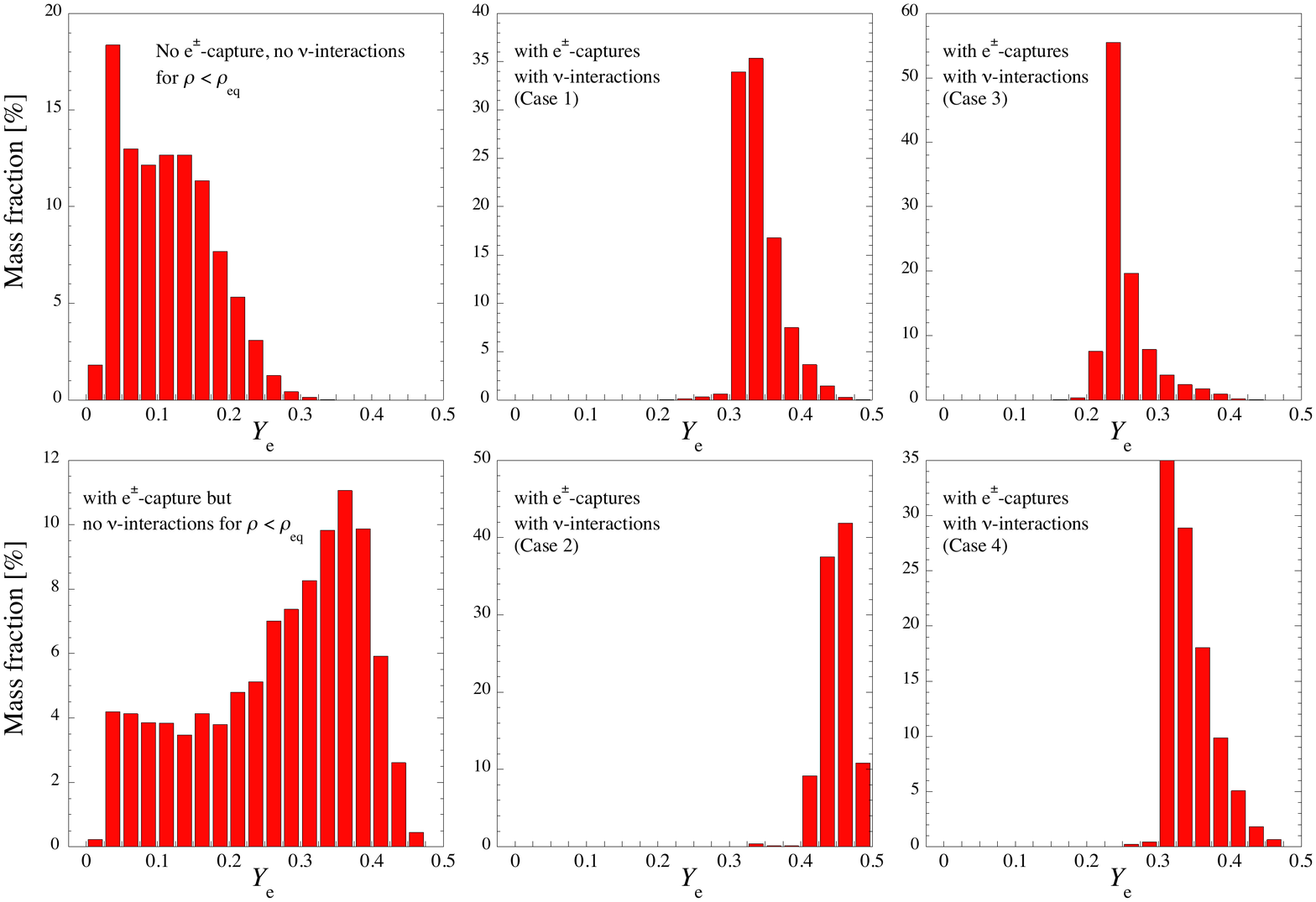}
  \caption{(Color online). Histograms of fractional mass distributions of the
    $4.9\times 10^{-3}$\Msun\ of matter ejected in our 1.35--1.35\Msun\ NS-NS merger model as functions 
    of $Y_e$ at density $\rho_{\rm net}$, assuming that no weak interactions of free nucleons
    take place below $\rho_\mathrm{eq}$ (upper left), that
    only electron and positron captures affect the $Y_e$ evolution for $\rho<\rho_\mathrm{eq}$ (lower left), 
    and including electron and positron captures as
    well as (anti)neutrino captures at $\rho < \rho_\mathrm{eq}$ 
    (Case~1: upper centre; Case~2: lower centre; Case~3: upper right; Case~4: lower right).
   }
\label{fig_histoye}
\end{figure*}

\subsection{$Y_e$ distributions}
\label{sec:yedistributions}

Six different cases are studied to estimate the impact of electron and positron captures as well as
electron (anti)neutrino absorption on the ejecta mass distribution as function of $Y_e$. 
In the first case, weak interactions of free nucleons are not 
allowed. In the second case, electron and positron captures are turned on, but not
(anti)neutrino absorptions. In the remaining four cases, both electron/positron captures as
well as (anti)neutrino captures are switched on, with the (anti)neutrino properties being 
defined by Cases 1--4. For these six cases, the resulting $Y_e$ distributions at density $\rho_{\rm net}$
are shown in Fig.~\ref{fig_histoye}.

In the first case without weak interactions of free nucleons below $\rho_\mathrm{eq}$ (upper left panel
of Fig.~\ref{fig_histoye}), the $Y_e$ distribution at $\rho_{\rm net}$ is identical to the one given at
the initial density $\rho_\mathrm{eq}$.  This case, however, differs from the standard case we considered
previously \citep{goriely11,bauswein13,goriely13,just15} in two aspects: First, we start our
calculations of weak interactions at density $\rho_\mathrm{eq}=10^{12}$\,g\,cm$^{-3}$ with $\beta$-equilibrium
distributions of electrons, positrons and neutrinos. This shifts and broadens the $Y_e$ distribution 
from previous values of $<$0.1 to a wider range between 0 and $\sim$0.3. Second, no temperature
post-processing is performed here. The initial temperatures of the trajectories are significantly higher
than those deduced from the post-processing applied in our previous studies.

As visible in Fig.~\ref{fig_histoye}, lower left panel, the $Y_e$ distribution at density 
$\rho_\mathrm{net}$ is significantly affected by $e^{\pm}$-captures between $\rho_\mathrm{eq}$
and $\rho_\mathrm{net}$, although some low-$Y_e$ ($\lsimeq 0.2$) ejecta are left. Dominant parts
of the ejecta are now found at $Y_e$ values between 0.3 and 0.4. When switching on neutrino
absorptions, for any of the Cases 1--4 the asymptotic values of $Y_e^{\nu\infty}$ are approached,
and further enhancement by the $\alpha$-effect produces peaks of the mass distributions
in a range of $Y_e$ values between 0.2 and 0.5. The peak values depend sensitively on the adopted
neutrino properties. Cases 1 and 4 lead to rather similar $Y_e$ distributions,
owing to the fact that the neutrino properties are broadly comparable.

\begin{figure}
\includegraphics[scale=0.3]{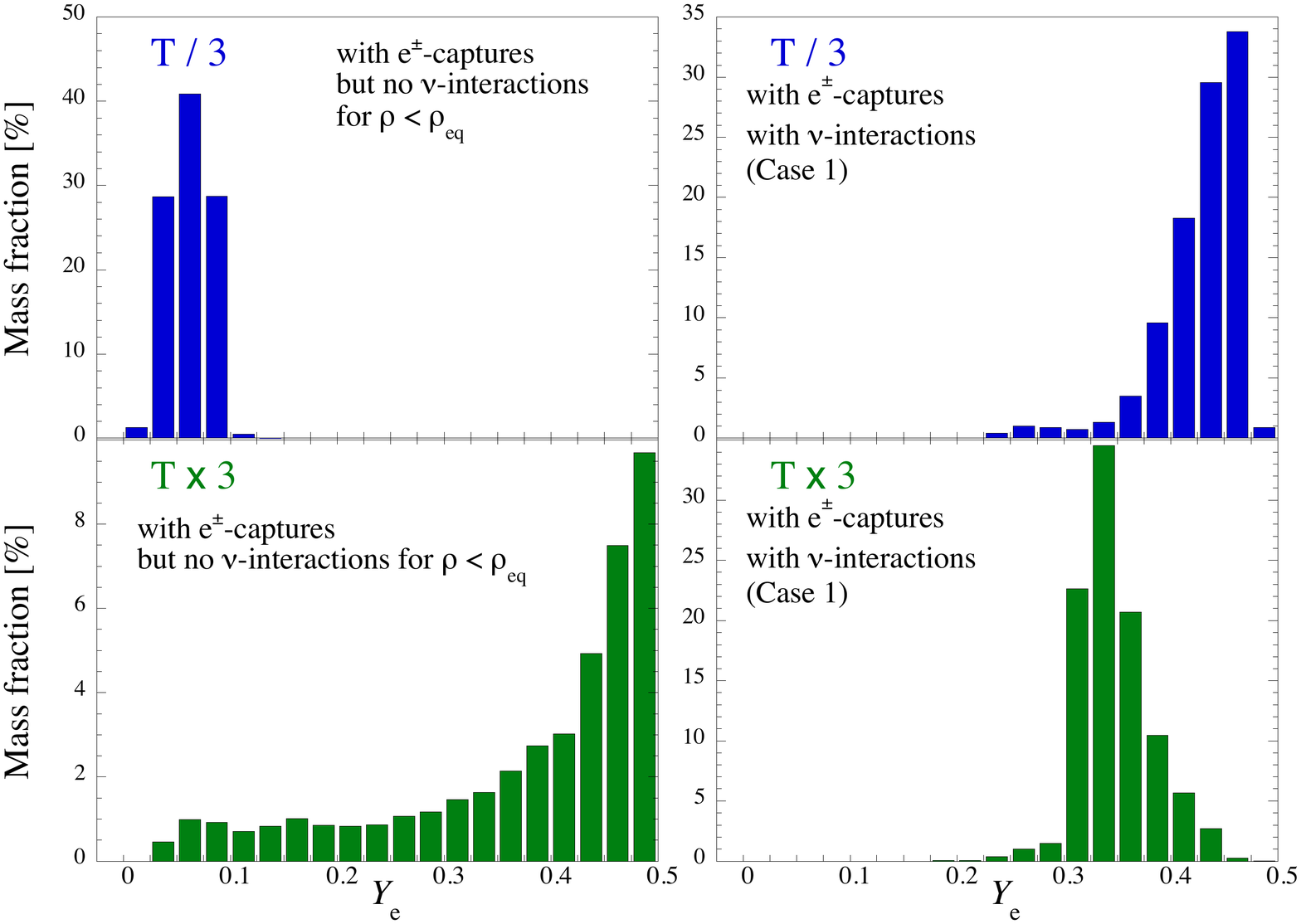} 
  \caption{(Color online). Histograms of fractional mass distributions as functions of $Y_e$ at
    density $\rho_{\rm net}$ for $e^\pm$ captures exclusively at $\rho<\rho_\mathrm{eq}$ (left) 
    and including electron, positron captures as well as (anti)neutrino captures during the
    evolution from $\rho_\mathrm{eq}$ to $\rho_{\rm net}$ (Case 1; right),
    when the temperatures are decreased (upper panels) or increased (lower panels)
    artificially by factors of~3.
    }
\label{fig_histoye_T}
\end{figure}

These distributions also depend sensitively on the temperature. 
When the temperature along the trajectory is
artificially increased or decreased by factors of 3, rather different results are obtained,
as shown in Fig.~\ref{fig_histoye_T}, using the (anti)neutrino properties of Case~1. 
As before we start the expansion evolution at $\rho_\mathrm{eq}$ with the $Y_e$ mass distribution
shown in the upper left panel of Fig.~\ref{fig_histoye}.
Reduced temperatures diminish the presence of positrons and thus favor electron captures compared
to positron captures. Moreover, lower temperatures also lead to a faster freeze-out of 
$e^\pm$ captures during the ejection of the mass elements. Without neutrino and antineutrino
absorptions, lower temperatures therefore tend to neutronize the ejecta for $\rho<\rho_\mathrm{eq}$
and the mass distribution becomes more narrow and is shifted to lower values
in the range of $0<Y_e\lsimeq 0.1$ (Fig.~\ref{fig_histoye_T},
upper left panel). Including neutrino and antineutrino absorptions, reduced temperatures have the
opposite effect in pushing the mass distribution to higher values of $Y_e$ (with a peak 
above 0.4) compared to the standard-temperature result for Case~1 in Fig.~\ref{fig_histoye} 
(upper right panel, with a peak of the distribution between $Y_e = 0.2$ and 0.3).
This behavior can be understood by the efficient recombination of free nucleons to $\alpha$ 
particles and heavy nuclei, which strengthens the heavy-nuclei ($\alpha$) effect so that
the peak of the distribution wanders to $Y_e\simeq 0.45 $. 
On the other side, increased temperatures reduce the electron degeneracy and thus 
allow for the presence of higher positron densities, thus enhancing positron captures
on neutrons. In addition, $e^\pm$ captures continue for a longer period of time along
the ejecta trajectories. Without (anti)neutrino absorption, these effects shift the 
$Y_e$ mass distribution from the initial one at $\rho_\mathrm{eq}$ (upper left panel
of Fig.~\ref{fig_histoye}) towards higher values of $Y_e$. This shift is stronger 
for more slowly expanding mass elements and weaker when the expansion is very
fast. Correspondingly, the mass distribution versus $Y_e$ at $\rho_\mathrm{net}$
is very broad and stretches from $\sim$0.03 up to a very  pronounced maximum close
to 0.5, because the $e^\pm$ capture equilibrium at high-entropy conditions favors
symmetric conditions with respect to neutrons and protons.
Taking into account (anti)neutrino absorption prevents this dramatic shift towards $Y_e\sim 0.5$,
because at large distances neutrino captures dominate $e^\pm$ absorptions and therefore
$Y_e$ asymptotes to values around $Y_e^{\nu\infty}$ ($\simeq 0.31$ for Case~1, lower right
panel of Fig.~\ref{fig_histoye_T}). Since the 
high temperatures favor nucleons and suppress the early formation of $\alpha$ particles
and heavier nuclei, the influence of the $\alpha$ effect is clearly weaker than in the
case of reduced temperatures (compare lower and upper right panels of 
Fig.~\ref{fig_histoye_T}).
As mentioned above, our calculations with varied temperatures are not consistent with the initial
$Y_e$ distributions used at a density of $\rho_\mathrm{eq}$, because these distributions are
calculated for the original ejecta temperatures provided by the hydrodynamic NS-NS merger model.
However, Fig.~\ref{fig_histoye_T} demonstrates that
asymptotic values determined by neutrino capture equilibrium and influenced by the
$\alpha$ (heavy-nuclei) effect are reached for most of the trajectories. The final $Y_e$
values can therefore be expected to mostly have lost the memory of the initial conditions at 
$\rho_\mathrm{eq}$. Correspondingly, when neutrino absorptions are included, the final mass 
distributions of $Y_e$ (at $\rho_\mathrm{net}$) are considerably more narrow than the
relatively broad distribution of initial $Y_e$ values before the expansion from 
$\rho_\mathrm{eq}$ to $\rho_\mathrm{net}$ (compare the right panels of
Fig.~\ref{fig_histoye_T} with the upper left panel of Fig.~\ref{fig_histoye}).

\section{r-process nucleosynthesis}
\label{sect_rpro}

\begin{figure*}
\includegraphics[scale=0.5]{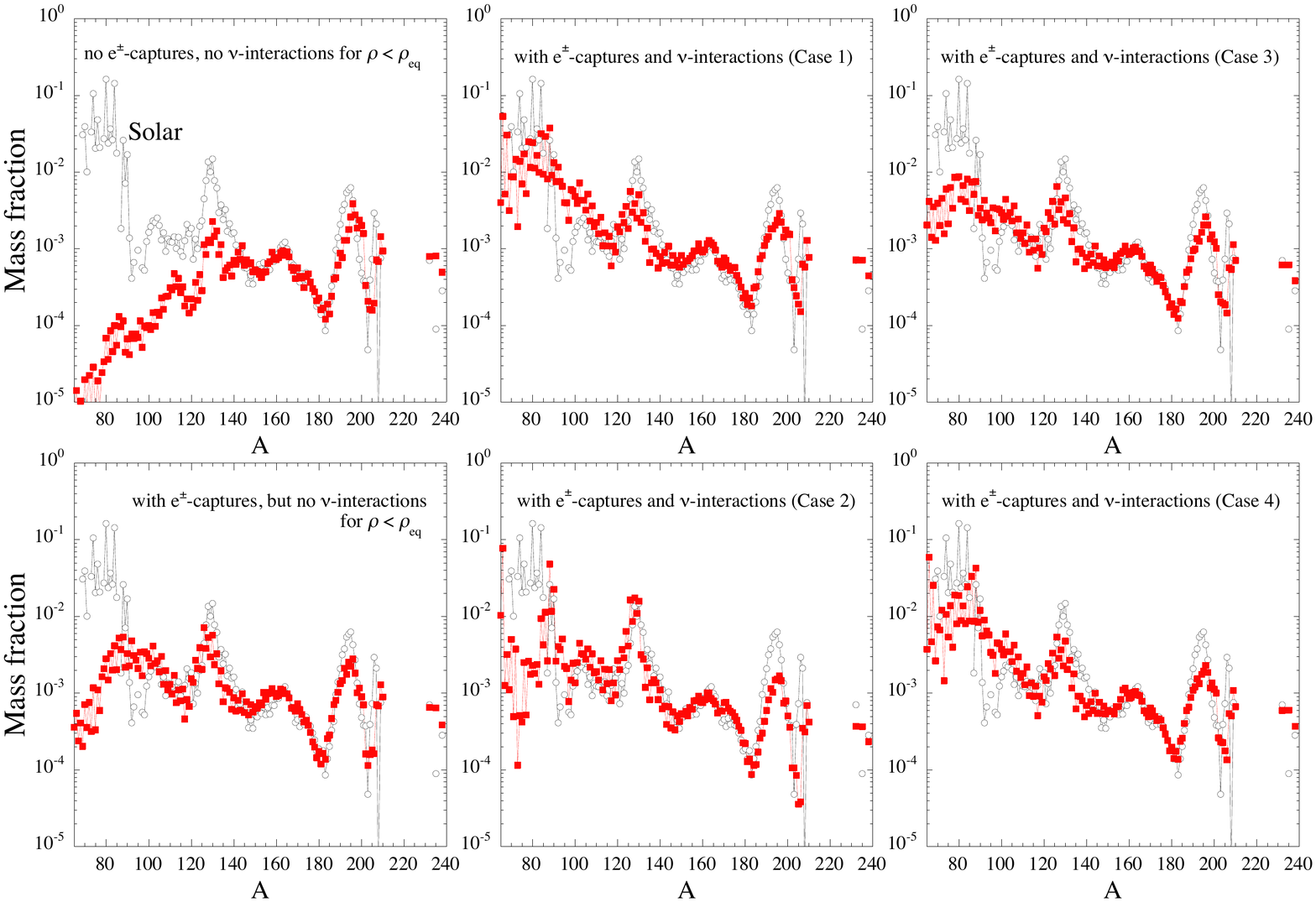}
  \caption{(Color online). Abundance distributions of the $4.9\times 10^{-3}$\Msun\ of material
    ejected in our 1.35--1.35\Msun\ NS-NS merger model, assuming that no weak interactions of free
    nucleons take place at $\rho<\rho_\mathrm{eq}$ (upper
    left), only electron and positron captures happen below $\rho_\mathrm{eq}$ (upper right), and
    electron and
    positron captures as well as (anti)neutrino captures affect the $Y_e$ evolution for 
    $\rho<\rho_\mathrm{eq}$ (Case~1: upper centre; Case~2: lower
    centre; Case~3: upper right; Case~4: lower right). The abundance distributions are
    normalized to the solar r-abundance distribution (open circles) in the rare-earth peak
    ($A=165$) region.}
\label{fig_rpro}
\end{figure*} 

At densities $\rho<\rho_{\rm net}$, the abundance evolution is followed by a full reaction network
\citep[for more details, see][]{goriely11,bauswein13,goriely13,just15}, which in contrast to
earlier studies includes now also the weak interactions of free
nucleons as detailed in Sect.~\ref{sect_weak}. 
The final abundance distributions are shown in Fig.~\ref{fig_rpro} for the same six
different cases discussed in Sect.~\ref{sec:yedistributions} and displayed in 
Fig.~\ref{fig_histoye}, namely for a case without weak interactions of free nucleons below
$\rho_\mathrm{eq}$, a case including $e^{\pm}$-captures but without (anti)neutrino absorption
reactions, and four cases where both $e^{\pm}$- and (anti)neutrino captures are taken into
account at $\rho<\rho_\mathrm{eq}$ (Cases 1--4 for the $\nu$ properties).

Because of its low-$Y_e$ distribution, the case neglecting $\beta$-interactions gives rise to 
an elemental abundance distribution similar to the one obtained in our previous studies
\citep{goriely11,bauswein13,goriely13,just15}. It is characterized by the production of essentially
only $A>140$ nuclei through several loops of fission recycling. When $e^\pm$-captures on free
nucleons are switched on, some low-$Y_e$ ($\lsimeq 0.2$) material can still lead to
nucleosynthesis with fission recycling and a significant production of the third r-process peak, but
the higher-$Y_e$ (0.3--0.4) matter can now also contribute to the production of
$90 \le A \le 140$ nuclei with a strong second $N=82$ peak.

\begin{figure}
\includegraphics[scale=0.3]{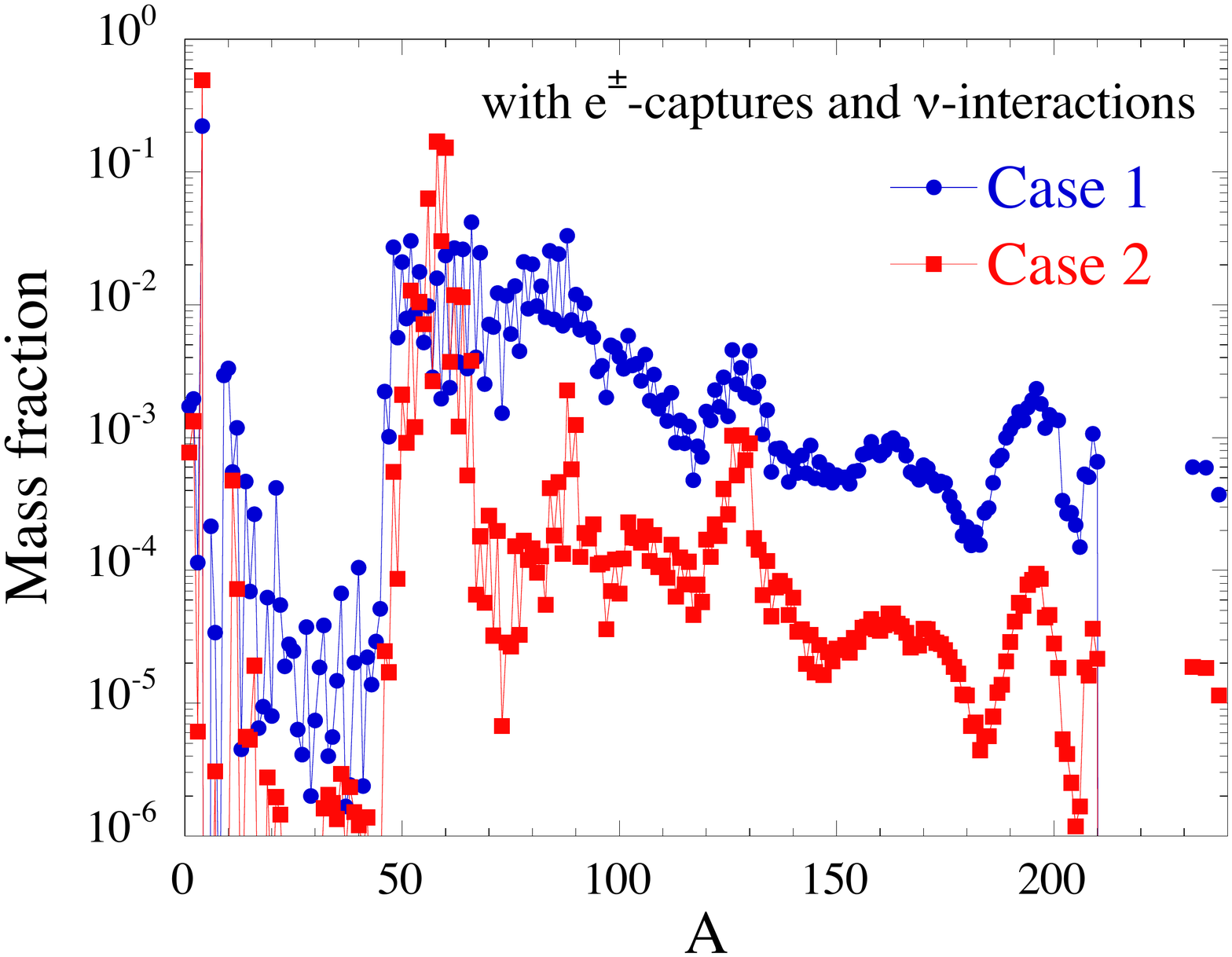}
  \caption{(Color online). Mass fraction as a function of the atomic mass for the matter ejected by our 1.35--1.35\Msun\
    NS merger model, including electron, positron as well as (anti)neutrino captures for Cases~1 and 2
    of the neutrino properties. 
       }
\label{fig_rpro_nu}
\end{figure}

When (anti)neutrino captures are included, too, nucleosynthesis with fission recycling does
not take place any longer, 
but the final abundance distribution still resembles the one in the solar system fairly well. A
significant amount of $50 \le A \le 90$ nuclei can now also be produced in addition to the $A>90$
r-nuclei, especially for (anti)neutrino properties corresponding to Case~2, as shown in
Fig.~\ref{fig_rpro_nu}. In this case, important element production around $^{60}$Ni is obtained,
originating from ejected mass elements with $Y_e \gsimeq 0.45$.

Without $\beta$-interactions below $\rho_\mathrm{eq}$, about 90\% of the ejected material is 
found to be r-process rich. If
$e^\pm$-captures on nucleons are switched on, the material is made of 76\% r-process nuclei, and when
(anti)neutrino captures are effective, too, we find that 45\% of the ejected matter is made of r-nuclei in
Case~1, only 1.9\% in Case~2, 67\% in Case~3 and 40\% in Case~4. In the last four cases, a significant
part of the material is made of $\alpha$-particles ($\sim 22$\% in Cases~1 and 4, 50\% in Case~2, and
17\% in Case~3), and the remaining part consists of $50 \le A \le 70$ nuclei (see Fig.~\ref{fig_rpro_nu}).
Despite the rather robust production of a solar-like distribution of r-nuclei, the absolute and relative
amounts of r-material vary strongly from case to case, sensitively depending on the neutrino exposure
of the ejecta as determined by the assumed properties of the neutrino emission. Neutrino properties
corresponding to Case~2 yield a total amount of ejected r-material that is significantly smaller
than for the other three sets of neutrino properties.

\begin{figure}
\includegraphics[scale=0.3]{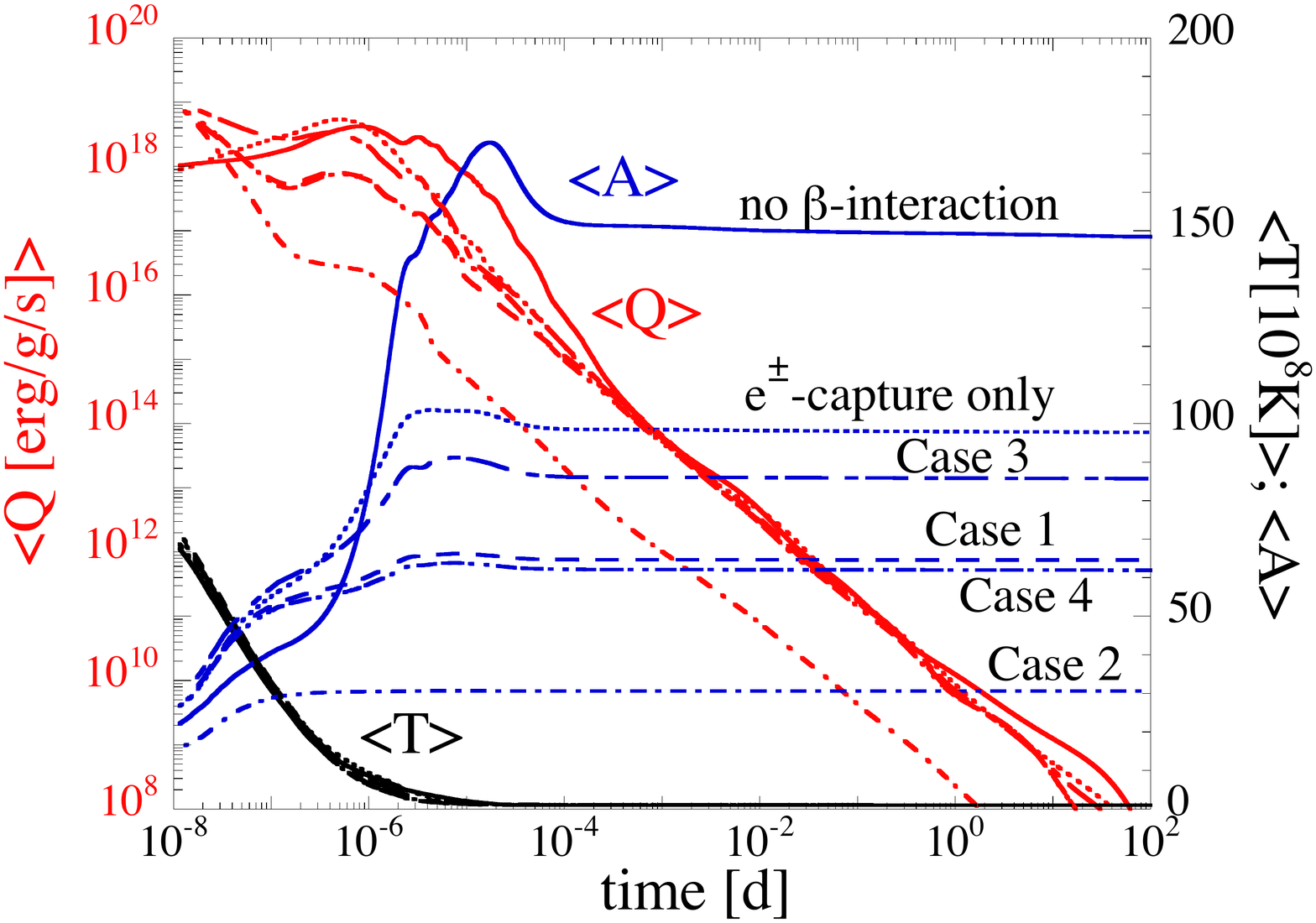}
\caption{(Color online). Time evolution of the total radioactive heating rate per unit mass,
$\langle Q\rangle$ (red), mass number $\langle A\rangle$ (blue), and temperature $\langle T\rangle$
(black), all mass-averaged over the ejecta, for the 1.35--1.35\Msun\ NS merger. The solid lines
correspond to the case without $\beta$-interactions of free nucleons at $\rho<\rho_\mathrm{eq}$, 
the dotted lines to the case
where only $e^\pm$-captures are taken into account, the dashed lines to the case where $e^\pm$ and
(anti)neutrino captures for neutrino properties according to Case~1 are included, the dash-dot 
lines for neutrinos according to Case~2, the long-dash-dot lines correspond to Case~3, and
the double-dash-dot lines to Case~4.}
\label{fig_QTA}
\end{figure}

The total, mass-averaged nuclear energy-release rate that is 
available for heating the ejecta per unit of mass, the average temperature of the ejecta,
and the average atomic mass number of the ejected abundance yields
are plotted as functions of time for our six studied cases in Fig.~\ref{fig_QTA}.
After some 10\,s and up to nearly one day, the energy release rates of all cases are fairly 
similar except for the treatment of neutrinos and $\beta$-interactions according to Case~2.
In this case,  the large electron fractions of the ejecta favor the production of light
elements and only a small amount ($\sim 2\%$) of the ejecta material consists of r-process nuclei
(Fig.~\ref{fig_rpro_nu}) and is therefore subject to fast $\beta$-decays. At late times, typically after one day, an additional source of energy is found from the $\alpha$-decay of long-lived heavy nuclei that can only be significantly  produced when no $\beta$-interaction of nucleons are included below $\rho_\mathrm{eq}$. 
Despite the considerable
differences of the nuclear energy-release rates at early times, the 
cooling evolution as measured by the mass-averaged temperatures is rather similar 
in all cases. 

The time evolution of the average nuclear
mass number of the ejected material clearly shows that only without $\beta$-interactions
of free nucleons a significant amount of fissile nuclei can be produced with
$\langle A \rangle$ reaching values up to 170, which is slightly below the value of about 200 
obtained when the $Y_e$ distribution of the inner crust of a cold NS is considered as initial state
\citep[see, in particular,][]{goriely11}. The sequence of 
decreasing values of $\langle A \rangle$ follows basically the hierarchy of
increasing values of the mean electron fraction, namely $\langle Y_e \rangle=0.11$ when 
$\beta$-interactions are ignored, $\langle Y_e \rangle=0.27$ when only $e^{\pm}$-captures
are taken into account, and 
$\langle Y_e \rangle=0.25$, 0.34, 0. 35, 0.45 in the Cases~3, 1, 4 and 2, respectively. The only
(slight) inversion is obtained for the calculation with only $e^{\pm}$-captures, because
in this case the $Y_e$ distribution is very wide (Fig.~\ref{fig_histoye}, lower left panel)
and the significant amounts of low-$Y_e$ ($\sim 0.05 - 0.15$) material contribute
to the production of heavy nuclei which increase the average nuclear mass number
$\langle A \rangle$.

\section{Summary and conclusions}
\label{sect_conc}

In this paper we reported the results of a parametric study to investigate
how $\beta$-interactions of free nucleons can affect the $Y_e$ evolution and mass
distribution in NS merger ejecta and the corresponding nucleosynthesis. To this
end we used the temperature-density trajectories of a large set of mass elements
representing the $\sim 5\times 10^{-3}\,M_\odot$ of matter ejected in a relativistic
merger simulation of a symmetric 1.35\Msun-1.35\Msun\ NS binary with the non-zero
temperature DD2 nuclear equation of state. Using the total lepton number provided
by our NS-NS merger simulation, we assume matter to be in $\beta$ equilibrium at
the temperature of the hydrodynamical model and a fiducial density of 
$\rho_\mathrm{eq}=10^{12}\,$g\,cm$^{-3}$. These conditions define the starting points
of our post-processing of the composition histories of the considered ejecta elements, 
for which we take into account electron and positron captures as well as electron
neutrino and antineutrino absorption on free neutrons and protons. 
Below a density $\rho_\mathrm{net}$, where the
temperature has decreased to $10^{10}$\,K (or the neutron drip density, if matter below
$\rho_\mathrm{eq}$ remains cooler than $10^{10}$\,K) 
the full network calculation is applied instead
of nuclear statistical equilibrium. Our description of weak interactions of free
nucleons includes weak magnetism and recoil corrections according to 
\citet{horowitz99} and \citet{pllumbi14}. Avoiding the complications of treating neutrino
transport, we simply use exemplary data from publications for prescribing the 
neutrino luminosities and mean energies needed to compute the neutrino absorption
rates in our parametric approach (``Cases~1--4''). 
This elementary prescription also accounts for the 
still large uncertainties of the model predictions for the neutrino emission and its
directional asymmetries in the generically three-dimensional merger scenario.

Our modeling strategy follows the spirit of previous, numerous parametric investigations
of nucleosynthesis in the neutrino-driven wind of newly formed neutron stars
in supernovae \citep[e.g.,][and references therein]{mclaughlin96,meyer98,arnould07,pllumbi14}
and of accretion tori around black holes as remnants of compact object mergers
\citep[e.g.,][]{surman05,surman08,wanajo12,caballero12}.
In particular, we track in detail the charged-current $\beta$-interactions of neutrinos with
free nucleons, which determine the evolution of the electron fraction outside of
the neutrino-trapping regime, and employ a full set of trajectories that characterizes the 
conditions in dynamical ejecta from the merging phase of a representative binary neutron star.
These conditions differ from proto-neutron star and accretion-torus winds not only 
concerning the range of entropies. 
The ejecta, especially, possess much faster expansion velocities,
which for the bulk of the matter can be 25--50\% of the speed of light, for some
fraction of the ejecta even faster, whereas neutrino-driven proto-neutron star
winds have typical velocities of 3--7\% of the speed of light 
\citep[e.g.,][]{arcones07}, neutrino-driven winds from massive neutron stars as relics
of NS mergers may achieve expansion velocities up to 8--10\% of the speed of light \citep{perego14},
and the main mass of neutrino-driven outflows from BH-accretion 
tori can reach 10--20\% of the speed of light \citep{just15}. 
The correspondingly shorter expansion time scales of dynamical merger ejecta
can enable a strong r-process even for moderately low $Y_e$.

Our calculations confirm recent results of \citet{wanajo14} that solar-like
r-process abundances are produced in dynamical NS merger ejecta even when 
$\beta$-reactions of free nucleons are taken into account and lead to a
significant increase of the average electron fraction. In contrast to 
previous works, where charged-current neutrino-nucleon interactions were
ignored, however, also nuclei with mass numbers $A<140$ are ejected in 
larger amounts. 

In detail, our results can be summarized by the following points:
\begin{itemize}
\item
Ignoring $\beta$-interactions at densities $\rho < \rho_\mathrm{eq}$, we
confirm our previous results of \citet{goriely11,bauswein13,goriely13,just15}
that almost exclusively r-nuclei in the regime $A\gsimeq 140$ are produced,
in spite of a moderate increase of the average ejecta $Y_e$ associated with the 
assumption of $\beta$-equilibrium at density $\rho_\mathrm{eq}$ instead of
our previous use of electron fractions of cold neutron star crust matter.
\item 
Positron and electron-neutrino captures, enhanced by weak magnetism corrections
(which increase the absorption cross section of $\nu_e$ and reduce that of 
$\bar\nu_e$) and supported by the $\alpha$ effect, lead to a shift of the 
average ejecta $Y_e$ towards higher values when matter expands downwards from
an initial density of $\rho_\mathrm{eq}=10^{12}$\,g\,cm$^{-3}$. 
\item
Captures of $e^\pm$ cause a wide spread of the $Y_e$ mass distribution,
reaching from values of $Y_e\ll 0.1$ up to 0.4--0.5. This reflects the wide 
range of thermodynamic conditions of the ejecta at $\rho_\mathrm{eq}$ with
ejecta trajectories that describe cool as well as hot conditions. 
Absorption processes of $\nu_e$ and $\bar\nu_e$
take over when the temperature of the expanding matter has fallen to 
low values where $e^\pm$ absorptions become ineffective. These (anti)neutrino
captures tend to push $Y_e$ towards the asymptotic value $Y_e^{\nu\infty}$
for $\nu_e$-$\bar\nu_e$
capture equilibrium. The corresponding $Y_e$ mass distributions (at 
$\rho_\mathrm{net}$, where the full network calculation was started) are
rather narrow in all cases, with a spread of $Y_e$ values of only 0.1--0.15.
The mean values of the distributions, however, vary between $\sim$0.25 and
$\sim$0.45, depending on the relative size of the $\nu_e$ and $\bar\nu_e$
luminosities. The latter are highly uncertain and variable and are sensitive
to the binary-parameter dependent merger dynamics, the nuclear equation of 
state, and the still not well determined directional asymmetries of the 
neutrino emission.
\item
In the presence of $\beta$-interactions of free nucleons the temperature
also plays an important role for the $Y_e$ evolution of the ejecta. 
It is another aspect of the hydrodynamical
merger models that depends on the system properties and the detailed ejection
dynamics (which differ between different ejecta components) and can be 
numerically problematic, because resolution and 
numerical/artificial viscosity can have an influence on the accuracy of
the determination of the thermal conditions. Lower temperatures reduce the
effects of $e^\pm$ captures but increase the importance of the $\alpha$
effect (thus pushing the average $Y_e$ closer to 0.5), whereas higher 
temperatures enhance $e^\pm$ captures but nevertheless have little influence 
on the final $Y_e$ distribution at $\rho_\mathrm{net}$, for which the
asymptotic value of $\nu_e$-$\bar\nu_e$ capture equilibrium is more relevant
than the initially fast $e^\pm$ captures.
\item
In all investigated model cases, the production of heavy r-process matter 
with strong second and third abundance peaks and a near-solar distribution 
in the rare-earth region is a robust outcome. However, in contrast to 
previous results where weak interactions of free nucleons were ignored,
also considerable amounts of matter are synthesized to $A<140$ nuclei.
The strength of the production of $A\lsimeq 90$--100 material is 
sensitive to the neutrino-emission properties that determine the 
(anti)neutrino absorption. In extreme cases where $Y_e$ gets close
to 0.5, significant amounts of iron-group nuclei ($A\sim 50$--60) can
be ejected. The relative fraction of heavy r-process matter (from the 
second peak upward) in the ejecta therefore varies dramatically between
the different investigated cases of neutrino-emission conditions and 
spans a range from more than 75\% down to just 2\%.
\end{itemize}

We emphasize that our parametric approach is highly simplified and
ignores important neutrino-transport effects like the exact spectral
distribution of the neutrino fluxes, the direction dependence of the
neutrino emission and corresponding precise radial dilution function,
and the time evolution of the neutrino emission relative to the
ejection time of the matter. Nevertheless,
our results, in support of those of \citet{wanajo14} and \citet{seki15}, 
have important consequences for the further exploration of the
nucleosynthesis connected to compact binary mergers and the discussion
of astrophysical implications.
In fact, they demand a major revision of the current picture of
r-process production in such events. 

In view of our results it is obvious that a proper treatment of the 
neutrino physics,
in particular of the neutrino irradiation of the ejected material, is
essential for making quantitative predictions of the elemental yields 
and especially of the total mass of r-process material that is thrown out
by the dynamical merger ejecta. Approximations like the ones used in our
study can be satisfactorily removed only when ultimately detailed, 
three-dimensional neutrino transport is consistently included in the 
hydrodynamical simulations.

Our study suggests that the relative contributions of matter with
$A\lsimeq 90$, $90\lsimeq A\lsimeq 140$ and $A\gsimeq 140$ are likely to
depend strongly on the binary properties and even the direction of mass ejection,
in addition to the equation of state dependence that was found for the 
electron-fraction distribution in the recent work of \citet{seki15}. 
Different from expectations so far, this means
that symmetric or nearly symmetric binary NS mergers could exhibit a 
significantly different ejecta composition than highly asymmetric NS-NS 
mergers and NS-BH mergers, in which the NS is disrupted before it can be
swallowed by the BH. In the last two cases 
the lower-mass component develops an extended tidal tail, from which
considerable amounts of cold, unshocked matter can be centrifugally 
ejected before the neutrino luminosities rise high and thus before
neutrino exposure of these ejecta plays an important role. In such a 
situation the ejecta will not only be expelled highly anisotropically but
will also carry a far dominant fraction of the mass in the form of
$A\gsimeq 140$ nuclei as predicted in previous studies
\citep[e.g.,][]{goriely11,bauswein13,goriely13,just15}.
In contrast, in symmetric or nearly symmetric NS mergers the contribution
of $A\lsimeq 140$ material will be higher. If the merger remnant collapses
to a BH on a millisecond time scale, neutrino exposure of the ejecta may
be avoided, but the broad $Y_e$ distribution caused by $e^\pm$ captures 
will allow for the production of $A>90$ nuclei with a strong second $N=82$ peak.
If the merger remnant remains transiently or permanently stable, neutrino
exposure of the dynamically expelled matter becomes important, enabling
a higher production of $A\lsimeq 90$ species.
For extreme cases of a luminous $\nu_e$ flux, the ejecta might then even
be dominated by iron-group nuclei including radioactive nickel isotopes.
In particular,
however, the exact composition and the relative fraction of high-mass
and low-mass species could depend on the direction of the mass ejection.
If most of the neutrino flux is emitted to the polar directions due to
the rotational deformation of the merger remnant, matter
expelled near the equatorial plane will receive less neutrino exposure in
addition to its potentially faster escape. This will allow for more
neutron-rich conditions close to the equator whereas polar ejecta
may contain more proton-rich contributions. Future, more complete merger
models with neutrino transport will have to clarify these possibilities.

Since the photon opacity, $\kappa$, of the expanding gaseous ejecta 
is strongly dependent on the presence of high-opacity, complex ions 
\citep[the lanthanides;][]{barnes13,kasen13,tanaka13,tanaka14}, the 
relative contribution of trans-iron elements to the ejecta will have
a severe impact on the peak luminosity,
$L_\mathrm{peak}\propto \kappa^{-1/2}$, peak time, 
$t_\mathrm{peak}\propto \kappa^{1/2}$, and the effective peak temperature, 
$T_\mathrm{peak}\propto \kappa^{-3/8}$, of the 
electromagnetic transient that is expected from the radioactively
heated, dynamical ejecta cloud 
\citep[``macronova'' or ``kilonova'',][]{lipaczynski98,kulkarni05,metzger10,roberts11,goriely11}.
The current picture of merger-type and remnant-type dependent redder 
(near-infrared) or bluer emission \citep{metzger14,perego14} and in particular of
the envisioned late-time infra-red radiation component from the dynamical
ejecta \citep{kasen14} might require revision or extension
in view of our results.

Finally, it is evident that the strong impact of neutrinos on the 
neutron-to-proton ratio and the nuclear composition of the dynamic
merger ejecta must move neutrino oscillations into the focus of interest.
It will be necessary to study the effects of collective neutrino oscillations
\citep[see, e.g.,][for a review]{duan10} with similiar intensity as this subject
currently receives in the context of the neutrino emission from newly born
neutron stars in supernovae.

\section*{Acknowledgments}
SG acknowledges financial support from FNRS (Belgium).
At Garching, this research was supported by the Max-Planck/Princeton Center for Plasma
Physics (MPPC) and by the Deutsche Forschungsgemeinschaft through the Cluster of Excellence EXC 153
``Origin and Structure of the Universe'' (http://www.universe-cluster.de). AB is a Marie Curie
Intra-European Fellow within the 7th European Community Framework Programme (IEF 331873). 
We are also grateful for computing time at the Rechenzentrum Garching (RZG).

\label{lastpage}
\end{document}